\documentclass[twocolumn,showpacs,preprintnumbers,amsmath,amssymb]{revtex4}%
\usepackage{amsfonts}
\usepackage{graphicx}
\usepackage{dcolumn}
\usepackage{bm}
\usepackage{amsmath}
\usepackage{amssymb}%
\begin{document}
\author{}

\bigskip%

\author{Hong Y. Ling$^1$, Han Pu$^2$ and Brian Seaman $^3$ }
\affiliation{$^1$Department of Physics, Rowan University,
Glassboro, NJ 08028-1700, USA \\ $^2$Department of Physics and
Astronomy, and Rice Quantum Institute, Rice University, Houston,
TX 77251-1892, USA
\\ $^3$JILA and Department of Physics, University of Colorado at Boulder, CO 80309-0440, USA }
\title
{Creating stable molecular condensate using a generalized Raman adiabatic passage scheme}
\begin{abstract}
We study the Feshbach resonance assisted stimulated adiabatic
passage of an effective coupling field for creating stable
molecules from atomic Bose condensate. By exploring the properties
of the coherent population trapping state, we show that, contrary
to the previous belief, mean-field shifts need not to limit the
conversion efficiency as long as  one chooses an adiabatic passage
route that compensates  the collision mean-field phase shifts and
avoids the dynamical unstable regime.
\end{abstract}
\date{\today}
\maketitle
%EndExpansion

Molecules offer a whole new dimension in the study of ultracold
atomic physics. In particular, resonant photo- or
magneto-association (Feshbach resonance \cite{Feshbach}) of cold
atoms to molecules represents a matter-wave analog of second
harmonic generation and has become a new paradigm of coupled
macroscopic quantum systems. Due to energy conservation, such
processes generally produce molecules in a vibrationally and/or
electronically excited quasibound level and hence are not
energetically stable. This does not seem to be a serious problem
in the case of fermionic atoms due to the suppression of molecular
decay by Pauli blocking. In fact, several groups have now
successfully Bose condensed so-formed molecules \cite{fbec}. In
contrast, although evidence of macroscopic coherence has been
observed in several experiments, excited molecules formed by a
pair of bosonic atoms have very short lifetime ($\lesssim$ a few
ms) and the atom-molecule conversion efficiency is limited to
$\lesssim 10 \%$ \cite{bomo,grimm}. Hence it is very important to
be able to create deeply-bound ground state molecules from atomic
Bose condensates with high efficiency, which is the focus of the
current work.

First proposed for photoassociating nondegenerate atoms into stable molecules
\cite{thermal}, stimulated Raman adiabatic passage (STIRAP) aided by Feshbach
resonance is considered a more efficient way of converting atomic condensates
into molecular ones \cite{kokk,mackie} than the bare STIRAP implementation of
photoassociation \cite{thermal,Mackie00}. In this scheme, the free atomic, the
quasibound and the ground molecular states form the three-level $\Lambda
$-system to which STIRAP can apply \cite{stirap}. The success of STIRAP relies
on the existence of the coherent population trapping (CPT) state
\cite{Alzetta76}. In a linear $\Lambda$-system, the CPT state exists when the
two-photon resonance condition is satisfied, hence STIRAP can be
straightforwardly implemented by appropriately choosing the laser frequencies.
In the case of condensate, collisions between particles give rise to nonlinear
mean-field shifts which dynamically changes when population is transferred
from atomic state to molecular state. This poses a serious problem for STIRAP
as collisions shift the system out of the two-photon resonance. For typical
experimental parameters, a conversion efficiency of only $\sim20\%$ is
predicted with a somewhat complicated laser sequence containing seven Raman
pulses \cite{kokk}. A possible remedy is to use a low-density atomic
condensate where the effect of collisions can be minimized \cite{juha}. This,
however, creates new problems as stronger coupling fields and longer time
scale are required.

The purpose of our work is to show that, despite of the presence
of the mean-field collisions, the \textquotedblleft
two-photon\textquotedblright\ or the CPT condition [See Eq.
(\ref{generalized two-photon resonance}) bellow] can be
\emph{dynamically maintained} and high atom-molecule conversion
rate can be achieved in condensates of typical density. The
nonlinear collisions, however, may induce dynamically unstable
regimes in the parameter space. It is crucial for the success of
STIRAP to avoid these unstable regimes when designing the route of
adiabatic passage.

\begin{figure}[h]
\begin{center}
\includegraphics[width=2.5in ] {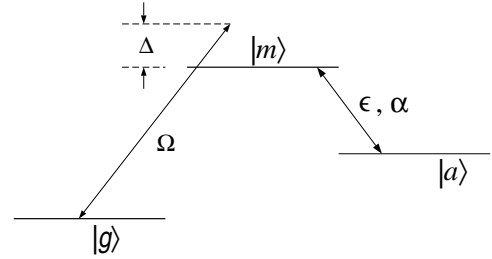}
\end{center}
\caption{The energy diagram of three-level atom-molecule system involving
free-quasibound-bound transitions.}%
\label{fig1}%
\end{figure}
%EndExpansion

Our model system is schematically sketched in Fig.~\ref{fig1},
where the relevant energy levels are denoted by $|a\rangle$ (free
atom), $|m\rangle$ (quasibound molecule) and $|g\rangle$ (ground
state molecule). Levels $|a\rangle$ and $|m\rangle$ are coupled by
a magnetic field through Feshbach resonance with coupling strength
$\alpha'$ and detuning $\epsilon$ ($\epsilon$ is experimentally
tunable via external magnetic field), while $|m\rangle$ and
$|g\rangle$ are coupled by a laser field with Rabi frequency and
detuning $\Omega$ and $\Delta$, respectively \cite{raman}. Without
loss of generality, we take both $\alpha'$ and $\Omega$ to be real
as their phase factors can be absorbed by a trivial global gauge
transformation of the field operators. In the interaction picture,
the Hamiltonian describing the system reads
\begin{align*}
\hat{H}  &  =\hbar\int d\mathbf{r}\,\left\{ \frac{1}{2} \sum_{i,j}%
\lambda_{ij}^{\prime}\hat{\Psi}_{i}^{\dag}(\mathbf{r})\hat{\Psi}_{j}^{\dag
}(\mathbf{r})\hat{\Psi}_{j}(\mathbf{r})\hat{\Psi}_{i}(\mathbf{r})\right. \\
&  +\epsilon\hat{\Psi}_{m}^{\dag}(\mathbf{r})\hat{\Psi}_{m}(\mathbf{r}%
)+\frac{\alpha^{\prime}}{2}\left[  \hat{\Psi}_{m}^{\dag}(\mathbf{r})\hat{\Psi
}_{a}(\mathbf{r})\hat{\Psi}_{a}(\mathbf{r})+h.c.\right] \\
&  +\left.  \left(  \Delta+\epsilon\right)  \hat{\Psi}_{g}^{\dag}%
(\mathbf{r})\hat{\Psi}_{g}(\mathbf{r})-\frac{\Omega}{2}\left[  \hat{\Psi}%
_{m}^{\dag}(\mathbf{r})\hat{\Psi}_{g}(\mathbf{r})+h.c.\right]  \right\}  ,
\end{align*}
where $\hat{\Psi}_{i}$ ($i=a$, $m$ and $g$) is the annihilation
field operator for state $|i\rangle$, the terms proportional to
$\lambda_{ij}^{\prime}$ represent two-body collisions with
$\lambda_{ii}^{\prime}$ $=4\pi\hbar a_{i}/m_{i}$ and
$\lambda_{ij}^{\prime}=\lambda_{ji}^{\prime}=2\pi\hbar
a_{ij}/\mu_{ij}$ for $i\neq j$ ($a_{i}$ and $a_{ij}$ are $s$-wave
scattering lengths, and $\mu_{ij}$ is the reduced mass between
states $i$ and $j$) characterizing the intra- and inter-state
interaction strengths, respectively. Here we consider a uniform
system and hence have dropped the usual kinetic and trapping
terms.

From the Hamiltonian we can easily derive the equations of motion
of the field operators. We adopt the standard mean-field treatment
to replace the field operators $\hat{\Psi}_{i}/\sqrt{n}$ with
$c$-number $\psi_{i}$, where $n$ is the density of the total
particle number. The mean-field approach ignores high order
quantum correlations, but is expected to be valid for systems with
a sufficiently large number of particles \cite{MFT}. The set of
the mean-field equations are
\begin{widetext}
\begin{subequations}
\label{dynamical equation}
\begin{eqnarray}
\frac{d\psi _{a}}{dt} &=&-i\left( \lambda _{a}\left| \psi
_{a}\right| ^{2}+\lambda _{am}\left| \psi _{m}\right| ^{2}+\lambda
_{ag}\left| \psi _{g}\right| ^{2}\right) \psi
_{a}-i\alpha \psi _{m}\psi _{a}^{\ast }, \\
\frac{d\psi _{m}}{dt} &=&-\gamma \psi _{m}-i\epsilon \psi
_{m}-i\left( \lambda _{m}\left| \psi _{m}\right| ^{2}+\lambda
_{am}\left| \psi _{a}\right| ^{2}+\lambda _{mg}\left| \psi
_{g}\right| ^{2}\right) \psi _{m}-i\frac{\alpha }{2}\psi _{a}^{2}+i\frac{%
\Omega }{2}\psi _{g}, \label{psim}\\
\frac{d\psi _{g}}{dt} &=&-i\left( \lambda _{g}\left| \psi _{g}\right|
^{2}+\lambda _{ag}\left| \psi _{a}\right| ^{2}+\lambda _{mg}\left| \psi
_{m}\right| ^{2}\right) \psi _{g}-i\left( \Delta +\epsilon \right) \psi
_{g}+i\frac{\Omega }{2}\psi _{m},
\end{eqnarray}
\end{subequations}
\end{widetext}
where $\lambda_{i}=\lambda_{ii}^{\prime}n$, $\lambda_{ij}=\lambda
_{ij}^{^{\prime}}n$, $\alpha=\alpha^{\prime}\sqrt{n}$ are the renormalized
quantities, and the term proportional to $\gamma$ in Eq.~(\ref{psim}) is
introduced phenomenologically to simulate the decay of the quasibound
molecules. \ Implied in Eqs.~(\ref{dynamical equation}) is the assumption that
this decay dominates all the other particle loss mechanisms. Hence,
Eqs.~(\ref{dynamical equation}) hold only in a time scale much shorter than
the characteristic times associated with other loss mechanisms, although the
time scale itself can be much longer than $\gamma^{-1}$.

The technique of STIRAP allows the system to evolve in a coherent
superposition of stable particle states ($|a\rangle$ and $|g\rangle$) over
time, effectively eliminating the loss of particles through highly unstable
states ($|m\rangle$). It is impossible to apply the STIRAP in a usual sense
directly to our model because our model contains nonlinear interaction terms.
Another difference with the conventional STIRAP is that while the coupling
strength of the $|m\rangle\leftrightarrow|g\rangle$ transition can be
controlled by the optical pulses, the coupling strength of the $|a\rangle
\leftrightarrow|m\rangle$ transition $\alpha$ is fixed not by optical means
but by the hyperfine interaction and is hence independent of time
\cite{Timmermans99}.

We now proceed to show that despite these important differences,
Eqs.~(\ref{dynamical equation}) support a CPT steady state with $\psi_{m}=0$.
To this end, let us first neglect the particle loss by taking $\gamma=0$ and
treat all the other parameters as time-independent. Then the total particle
number is conserved, i.e., $\left\vert \psi_{a}\right\vert ^{2}+2\left\vert
\psi_{m}\right\vert ^{2}+2\left\vert \psi_{g}\right\vert ^{2}=1, $ so that we
can introduce the atomic chemical potential $\mu$ through the following
steady-state ansatz
\begin{equation}
\psi_{a}=\left\vert \psi_{a}\right\vert e^{i\theta_{a}}e^{-i\mu t}%
,\;\;\psi_{m,g}=\left\vert \psi_{m,g}\right\vert e^{i\theta_{m,g}}e^{-i2\mu
t}. \label{decom}%
\end{equation}
Putting (\ref{decom}) into (\ref{dynamical equation}) and taking the time
derivative to be zero, one finds that the following CPT solution exists:
\begin{subequations}
\label{cpt population distribution}%
\begin{align}
\left\vert \psi_{a}^{0}\right\vert ^{2}  &  =\frac{2}{\sqrt{1+8\left(
\alpha/\Omega\right)  ^{2}}+1}=1-2\left\vert \psi_{g}^{0}\right\vert ^{2},\\
\left\vert \psi_{m}^{0}\right\vert ^{2}  &  =0,
\end{align}
with $\theta_{g} = 2 \theta_{a}$ and the corresponding chemical potential
$\mu=\lambda_{a}\left\vert \psi_{a}^{0}\right\vert ^{2}+\lambda_{ag}\left\vert
\psi_{g}^{0}\right\vert ^{2}$, under the condition
\end{subequations}
\begin{equation}
\Delta=-\epsilon+\left(  2\lambda_{ag}-\lambda_{g}\right)  \left\vert \psi
_{g}^{0}\right\vert ^{2}+\left(  2\lambda_{a}-\lambda_{ag}\right)  \left\vert
\psi_{a}^{0}\right\vert ^{2}. \label{generalized two-photon resonance}%
\end{equation}

Equation~(\ref{generalized two-photon resonance}) is the
generalized \textquotedblleft two-photon\textquotedblright\
resonance condition which incorporates the nonlinear collisional
phase shifts. This CPT solution is thus a generalization of the
one found in the collisionless limit \cite{Mackie00}.

Two crucial conclusions regarding the utility of this CPT state in the
adiabatic atom-molecule conversion can be arrived by observing
(\ref{cpt population distribution}) and
(\ref{generalized two-photon resonance}): i) Population is concentrated in
states $|a\rangle$ and $|g\rangle$ under the respective limit $\alpha
/\Omega\rightarrow0$ and $\alpha/\Omega\rightarrow\infty$, which facilitates
adiabatic coherent population transfer between the atoms and the ground state
molecules by tuning the optical Rabi frequency $\Omega$ ($\alpha$, as we
mentioned, is regarded as fixed in time); ii) Since the temporal dependence of
$\Omega$ uniquely determines the time evolution of the population, we can
design \emph{a priori} the temporal evolution of the laser detuning $\Delta$
in accordance of $\Omega(t)$ such that the resonance condition
(\ref{generalized two-photon resonance}) is maintained dynamically at any
time. \ (A similar idea can be found in a study of vortex coupler
\cite{Marzlin97}. ) \ In so doing, we eliminate the need to minimize the
effect of collisions by reducing the density of the system \cite{juha} or by
other means \cite{Drummond02}.

\begin{figure}[h]
\begin{center}
\includegraphics[width=2.3in]{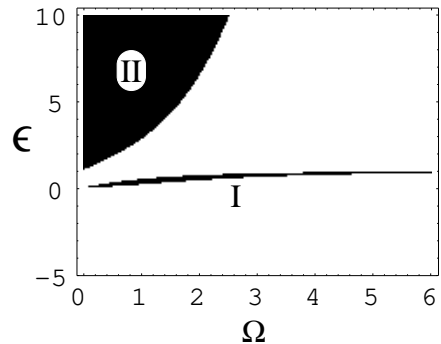}
\end{center}
\caption{ Stability diagram in $\left(  \Omega,\epsilon\right)  $-space.
$\Omega$ and $\epsilon$ are in units of $\alpha$. Other parameters:
$\lambda_{a}=0.625$, $\lambda_{m}=\lambda_{g}=0.1875$, $\lambda_{am}%
=\lambda_{ag}=\lambda_{mg}=0.1875$, all in units of $\alpha$. }%
\label{fig2}%
\end{figure}

The existence of the CPT state, however, does not guarantee that this state
can be followed adiabatically. We have to study the stability properties of
the state. This has been neglected in all the previous studies on this
problem. For this purpose, we adopt the linear perturbation approach, adding
small fluctuations (in both amplitudes and phases) to the steady-state CPT
solutions and linearizing the equations of motion. The excitation frequencies
of the system other than the zero frequency Goldstone mode can be found
analytically as
\begin{equation}
\omega=\pm\sqrt{\left(  b\pm\sqrt{b^{2}-4c}\right)  /2}, \label{omega^4}%
\end{equation}
where
\begin{align*}
b  &  =\Omega^{2}/2+2\alpha^{2}\left\vert \psi_{a}^{0}\right\vert ^{2}+A^{2},\\
c  &  =\left(  \Omega^{2}/4+\alpha^{2}\left\vert
\psi_{a}^{0}\right\vert ^{2}\right)
^{2}+\frac{\alpha^{2}}{2}\left\vert \psi_{a}^{0}\right\vert
^{4}\left(  \lambda_{g}-4\lambda_{ag}+4\lambda_{a}\right)A  ,
\end{align*}
with $A=\lambda|\psi_{a}^{0}| ^{2}+\lambda_{ag}-0.5\lambda
_{mg}-\epsilon$ and
$\lambda=2\lambda_{a}-\lambda_{ag}-\lambda_{am}+0.5\lambda_{mg}$.
When $\omega$ becomes complex, the corresponding CPT state is
\emph{dynamically unstable}. Hence the unstable regime is given by
either $c<0$ or $c>b^{2}/4$. The instability here is caused by
collisions as it can be easily seen that in the absence of
collisions, $\omega$ is always real. The typical results from the
stability analysis based on the parameters of our interest are
summarized in Fig.~\ref{fig2}, where the $\left(
\Omega,\epsilon\right)  $-space is divided into the stable (white)
and the unstable (dark) regions.

In our calculations, we have taken the parameters for $^{23}$Na. The $s$-wave
scattering length for sodium atom is well-known and is around $a_{a}=3.4$ nm
\cite{abeelen99} which yields $\lambda_{aa}^{\prime} =1.18\times10^{-16}$
m$^{3}$ s$^{-1}$. We choose $\alpha^{\prime}$ =$4.22\times10^{-6}$ m$^{3/2}%
$s$^{-1}$, corresponding to the atom-molecule coupling strength for the sodium
Feshbach resonance at a magnetic field strength of $85.3$ mT \cite{seaman03}.
We take the condensate density $n$ to be $5\times10^{20}$ m$^{-3}$. This gives
rise to $\alpha=9.436\times10^{4}$ s$^{-1}$, and $\lambda_{a}=5.9\times10^{4}$
s$^{-1}=0.625\alpha$. There are so far no good estimates on molecular
scattering lengths. So we take the collisional coefficients involving the
molecular levels to have the same magnitudes as $\lambda_{a}$. Generally
speaking, there are two unstable regions: Region I is thin along $\epsilon
$-dimension and centers at $\epsilon=\lambda_{ag}-\lambda_{mg}/2$ in the small
$\Omega$ limit and at $\epsilon=2\lambda_{a}-\lambda_{am}$ in the large
$\Omega$ limit; Region II occurs at small $\Omega$ and lies either above or
below Region I in $\epsilon$-direction depending on whether $\lambda
_{g}-4\lambda_{ag}+4\lambda_{a}$ is positive or negative. We will show that it
is very important to avoid these unstable regions in order to convert atoms
into stable molecules.

%EndExpansion

\begin{figure}[ptb]
\begin{center}
\includegraphics[width=2.5in ] {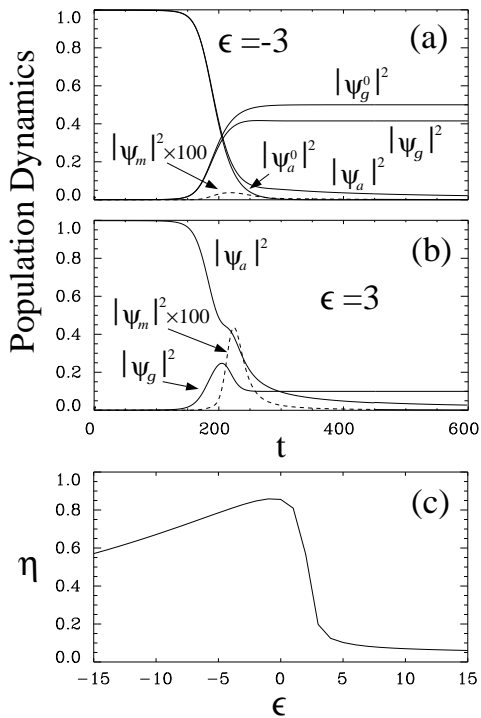}
\end{center}
\caption{(a) and (b): population as functions of time for $\epsilon=-3$ and 3,
respectively. Also plot in (a) are the CPT solutions of
Eq.~(\ref{cpt population distribution}). Here $\gamma=1$, $\Omega_{\max}=40$,
$\tau=40$, $t_{0}=120$, and all the other parameters are the same as in
Fig.~\ref{fig2}. Time is in units of $1/\alpha$, all other quantities are in
units of $\alpha$. (c) Conversion efficiency $\eta=2|\psi_{g}(t=\infty)|^{2}$
as a function of $\epsilon$.}%
\label{fig3}%
\end{figure}

Samples of our results on atom-molecule conversion are displayed in
Fig.~\ref{fig3}, where we have numerically solved the full set of
Eqs.~(\ref{dynamical equation}) including the loss term with a time-varying
Rabi frequency given by
\begin{equation}
\Omega\left(  t\right)  =\frac{\Omega_{\max}}{2}\left[  1-\tanh\left(
\frac{t-t_{0}}{\tau}\right)  \right]  , \label{applied field}%
\end{equation}
and $\Delta$ changes accordingly as determined by
(\ref{generalized two-photon resonance}). Although in principle
the Feshbach detuning $\epsilon$ can be varied temporally to
optimize the conversion efficiency, we find it not quite
necessary. Fixing $\epsilon$ at a finite value not only simplifies
the experimental procedure, more importantly it also allows us to
avoid strong condensate losses near the exact Feshbach resonance
($\epsilon=0$) \cite{loss}. An experimental implementation of our
scheme may thus take the following steps. First, turn on the
coupling field and fix its amplitude at $\Omega_{\max}$ while the
Feshbach transition is far-off resonance. Next, we bring the
Feshbach transition suddenly into near resonance. This is followed
by the adiabatic ramp of the amplitude and the frequency of the
coupling field according to Eq.~(\ref{applied field}) and
Eq.~(\ref{generalized two-photon resonance}), respectively. This
implementation represents a counterintuitive sequence as the
optical field coupling the initially empty state $\left\vert
g\right\rangle $ and $|m \rangle$ is turned on before the
interaction on the transition involving the initially populated
state $\left\vert a\right\rangle $.

%EndExpansion

Figure \ref{fig3}(a) shows the time evolution of the population $|\psi
_{a,g,m}(t)|^{2}$. Here $\epsilon$ is chosen such that the system remains in
the stable regime (see Fig.~\ref{fig2}). Also plotted in the figure is the
analytical CPT solutions of Eqs.~\ref{cpt population distribution}. As it can
be seen, the exact population dynamics follows closely the prediction of the
CPT solutions. The small discrepancies at later times can be attributed to the
fact that the system can not maintain adiabaticity completely \cite{adia}. The
final population in the stable molecular state is $|\psi_{g}(\infty
)|^{2}=0.415$, corresponding to a conversion efficiency of $\eta=2|\psi
_{g}(\infty)|^{2}=83\%$. The conversion efficiency can be further improved if
the optical field is ramped down more slowly. In contrast, in the dynamics
depicted in Fig.~\ref{fig3}(b), initially the populations follow the CPT
solutions, but significant deviation starts to occur at about $t=200/\alpha$
when the system enters the unstable regime. The final conversion efficiency is
less than 20$\%$, which cannot be improved by using a more slowly varying
optical field as one does in the stable regime. Figure~\ref{fig3}(c)
summarizes the conversion efficiencies for various $\epsilon$. The marked
asymmetry between positive and negative $\epsilon$ is a dramatic manifestation
of the dynamical instability.

In conclusion, we have presented here an efficient method to convert atomic
condensate into stable molecules using a single optical pulse without the need
to sweep through the Feshbach resonance. Our method generalizes the
conventional STIRAP scheme in that the laser is frequency chirped to
compensate for the mean-field shift arising from the particle collisions, so
that, at least in principle, the \textquotedblleft
two-photon\textquotedblright\ resonance is maintained at all time. We also
point out that the presence of particle collisions also gives rise to
potential dynamically unstable regimes in parameter space. A high conversion
efficiency can be reached only if one designs the STIRAP route such that these
unstable regimes are avoided.

Note that we do not expect the presence of the trap will change
our results significantly so long as the trapping frequencies are
made sufficiently small.  An estimate in the Thomas-Fermi limit
and for a spherical trap of frequency $\omega _{0}=2\pi \times 41$
Hz indicates that more than 90\% of a total of 5$\times 10^{6}$
sodium atoms will occupy the central part of the trap where the
variation of the mean-field shift due to density inhomogeneity is
less than $\alpha /10$. A numerical calculation based on the same
parameters as in Fig.~\ref{fig3} but with the two-photon resonance
condition violated by an amount of $\alpha/10$ indicates that a
conversion efficiency as high as 76 \% is still attainable.

Finally we want to mention that although we specifically studied a
Feshbach resonance situation, due to the formal resemblance of the
governing equations, our method can be straightforwardly applied
to a photoassociation situation. In fact, we solved Eqs.~(3) of
Ref.~\cite{juha} with our laser frequency modulation scheme and
obtained a final stable molecular population of 0.42 at density
$n=4.3\times10^{14}$cm$^{-3}$ using the parameters for their Fig.
2(c). Compared to 0.08 obtained in Ref.~\cite{juha}, the advantage
of our method is quite obvious.

HYL acknowledges the support from the US National Science Foundation under
Grant No. PHY-0307359 and HP is supported by Rice University.\ \ %


\begin{thebibliography}{1}
\bibitem{Feshbach}
H. Feshbach, {\em Theoretical Nuclear Physics} (Wiley, New York,
1992); E. Tiesinga, B.J. Verhaar, and H.T.C. Stoof, Phys. Rev A
{\bf 47}, 4114 (1993); E. Tiesinga {\em et al.}, {\em ibid.} {\bf
46}, R1167 (1992).
\bibitem{fbec}M. Greiner, C. Regal and D. S. Jin, Nature {\bf426}%
, 537 (2003); S. Jochim {\em et al.}, Science {\bf302}, 2101
(2003); M. W. Zwierlein {\em et al.}, Phys. Rev. Lett. {\bf91},
250401 (2003); T. Bourdel {\em et al.}, cond-mat/0403091.
\bibitem{bomo}E. A. Donley {\em et al.}, Nature (London) {\bf
417}, 529 (2002); J. Herbig {\em et al.}, Science {\bf 301}, 1510
(2003); K. Xu {\em et al.}, Phys. Rev. Lett. {\bf 91}, 210402
(2003); S. D\"{u}rr {\em et al.}, Phys. Rev. Lett. {\bf 92},
020406 (2004).

\bibitem{grimm}One notable exception is the recent work by the
group of R. Grimm (M. Mark {\em et al.}, cond-mat/0409731) who
emloyed a switching scheme to convert over $30 \%$ of Cs
condensate into weakly bound Cs$_2$ molecules. 

\bibitem{kokk}S.J.J.M.F. Kokkelmans, H.M.J. Vissers and B. J.
Verhaar, Phys. Rev. A {\bf63}, 031601 (2001).
\bibitem{mackie} M. Mackie, Phys. Rev. A {\bf66}, 043613 (2002).
\bibitem{thermal}A. Vardi {\em et al.}, J. Chem. Phys. {\bf107},
6166 (1997).
\bibitem{Mackie00}
M. Mackie, R. Kowalski, and J. Javanainen, Phys. Rew. Lett. {\bf84}, 2000.
\bibitem{Alzetta76}
G. Alzetta {\em et al.}, Nuovo Cimento B {\bf 36}, 5 (1976); G.
Alzetta, L. Moi and G. Orriols, {\bf52}, 205 (1979).
\bibitem{stirap}K. Bergmann, H. Theuer and B. W. Shore, Rev. Mod.
Phys. {\bf70}, 1003 (1998).
\bibitem{juha}M. Mackie {\em et al.}, Phys. Rev. A {\bf 70},
013614 (2004).
\bibitem{raman}In the case that the transition between $|m
\rangle$ and $|g \rangle$ is dipole forbidden, the coupling has to
be provided by a pair of Raman lasers via an additional
off-resonant intermediate molecular level. Then the Rabi frequency
and detuning should be regarded as the effective two-photon Raman
coupling strength and detuning.

\bibitem{MFT}A.S. Parkins and D. F. Walls, Phys. Rep. {\bf 303}, 1 (1998).
\bibitem{Timmermans99}
E. Timmermans {\em et al.}, Phys. Rep. {\bf 315}, 199 (1999).
\bibitem{Marzlin97}
K.-P. Marzlin, W. Zhang and E. M. Wright, Phys. Rev. Lett., {\bf
79}, 4728 (1997).
\bibitem{Drummond02}
P. D. Drummond {\em et al.}, Phys. Rev. A {\bf 65}, 063619 (2002).
\bibitem{abeelen99}
F. A. van Abeelen, and B. J. Verhaar, Phys. Rev. A {\bf 59}, 578
(1999).
\bibitem{seaman03}
B. Seaman and H. Ling, Opt. Commun. {\bf226}, 267 (2003).
\bibitem{loss}
J. Stenger {\em et al.}, Phys. Rev. Lett. {\bf82}, 2422 (1999).
\bibitem{adia}We plan to address the issue of adiabaticity in a
future publication.
\end{thebibliography}
\end{document}